\documentclass[12pt,floatfix,showpacs,showkeys,aip,reprint,apl]{revtex4-1}
\usepackage[utf8x]{inputenc}
\usepackage[british]{betababel} 
\usepackage{amsmath} 
\usepackage{booktabs} 
\usepackage{pgfplots}
\usepackage[separate-uncertainty = true,multi-part-units=single,per-mode=symbol-or-fraction]{siunitx} 
\usepackage{tikz}
\usepackage{xcolor}
\definecolor{darkblue}{cmyk}{1,0.94,0.24,0.19}
\definecolor{darkred}{cmyk}{0.31,0.95,0.72,0.31}
\definecolor{darkyellow}{cmyk}{0.34,0.38,1,0.06}

\date{\today}

\begin{document}
\author{Philipp Krüger}
\affiliation{KIT -- Karlsruhe Institute of Technology, Hermann-von-Helmholtz-Platz 1, 76344 Eggenstein-Leopoldshafen, Germany}
\author{Francesco Grilli}
\author{Michal Vojenčiak}
\author{Victor M.R. Zermeño}
\author{Eduard Demencik}
\affiliation{KIT -- Karlsruhe Institute of Technology, Hermann-von-Helmholtz-Platz 1, 76344 Eggenstein-Leopoldshafen, Germany}
\author{Stefania Farinon}
\affiliation{Istituto Nazionale di Fisica Nucleare, via Dodecaneso 33, 16146 Genova, Italy}
\title[SC/FM heterostructures: potential for reduced hysteretic losses]{Superconductor/ferromagnet heterostructures exhibit potential for significant reduction of hysteretic losses}
\email{philipp.krueger@kit.edu}

\begin{abstract}
First experimental observations of the ferromagnetic shielding effect in high-$\text{T}_{\text{c}}$ superconducting coated conductors were carried out. Experimental results were compared to simulations calling upon finite-element calculations based on the H-formulation of Maxwell equations to model superconducting strips with ferromagnetic shields. Samples of copper-stabilized coated conductors were electroplated with nickel shields and afterwards characterized. Both externally applied oscillating transverse magnetic fields as well as transport currents were studied. Having observed promising gains with respect to the reduction of ac losses in both cases, we further investigated the potential of ferromagnetic shielding. The numerical model was able to reproduce and also predict experimental results very well and will serve as an indispensable tool to determine the potential of soft ferromagnetic materials to significantly reduce hysteretic losses.
\end{abstract}

\maketitle

\sisetup{scientific-notation=engineering,number-unit-product=\text{~},exponent-product=\times,output-product=\times}


With hysteretic losses being extremely important to superconducting systems due to the requirement for cryogenic temperature regimes in which to operate, reducing energy dissipation is crucial. Presently, the most promising high-$\text{T}_{\text{c}}$ superconductors are \textit{RE}BCO (Rare Earth Barium Copper Oxide) coated conductors, also known as second generation tapes. These consist of a thin film of superconducting material deposited upon a buffered metallic substrate~\cite{Malozemoff2012}. Owing to the unique geometry of these tapes with very large aspect ratios, specialized solutions have to be devised in order to reduce their hysteretic losses. Apart from striating the tapes longitudinally~\cite{Carr1999,Cobb2002,Amemiya2004,Terzieva2011}, ferromagnetic constituents have been proposed as means to reduce hysteretic losses in superconductors in the past~\cite{Glowacki2000,Gomory2007,Gomory2009,Safran2010,Genenko2009,Gomory2010,Genenko2011a}. However, of the publications pertaining to coated conductors, none were able to supplement their simulations with reasonably complying measurements or only partly so~\cite{Genenko2009,Gomory2010,Genenko2011a}. Those exploiting magnetostatic-electrostatic analogues were unable to account for losses in the ferromagnetic domains~\cite{Genenko2009,Genenko2011a}.

Throughout the article, we understand the samples to be pieces cut from commercially available Superpower tape clad with \SI{20}{\micro\metre} of copper and a width $a$ of \SI{12}{\milli\metre}. The superconducting layer itself has a thickness of $d=\SI{1}{\micro\metre}$. In order to characterize the superconductor as completely as possible, we measured the critical current of both pristine as well as ferromagnetically shielded conductors. The latter were prepared by galvanization of the coated conductors in a Wurtz bath. To control the size of the area to be coated, self-adhesive plastic foil was used that could easily be removed later-on. For the measurements of hysteretic losses, two methods were employed: the calibration free magnetisation measurement with a differentiating two coil system~\cite{Souc2005} in the case of applied oscillating magnetic fields and the lock-in amplified integrating method~\cite{Hughes1994} in the self field case with applied currents. We only used lock-in amplifiers for reference measurements to compare against though. The actual measurements were done using high-speed data acquisition units that allowed for very precise measurements of current and voltage while maintaining phase alignment.

We measured the DC critical current $I_{\text{c}}$ of our samples before (\SI{290(10)}{\ampere}) and after coating them with nickel (\SI{284(10)}{\ampere}). Having not investigated changes in $I_{\text{c}}$ particularly, we are unable to state with final authority whether the critical current may be positively affected by the presence of ferromagnetic shielding as predicted in~\cite{Genenko1999,Genenko2000,Genenko2009}. Within the measurement accuracy, we did not observe a trend that would support this theory. If prepared carefully, the coating process does not significantly diminish critical currents with differences in $I_{\text{c}}$ before and after coating for all samples not being larger than \SI{5}{\percent}. In one instance, we even found an increase in critical current. In an earlier iteration of this process, we worked with coated conductors sporting only a thin silver shunt instead of being clad with copper. This silver shunt and the superconducting layer underneath were adversely affected by the acidic environment of the Wurtz bath used in the electroplating process to a degree that lowered critical current values to as low as \SI{150}{\ampere}, a decrease by \SI{50}{\percent}.

The numerical model we use to validate our measurements is based on the well-established H-formulation~\cite{Hong2006,Brambilla2007}, adapted to account for magnetic materials~\cite{Nguyen2010a}. In the case of ferromagnetically shielded samples, we are simulating a ferromagnetic nickel coat of \SI{20}{\micro\metre} thickness that covers \SI{25}{\percent} of the superconducting strip from each side (compare the sketch in fig.~1), resulting in \SI{50}{\percent} total coverage. The actual coat however is slightly thicker at the edges and shows pitting due to hydrogen bubbles clinging to the surface during the electroplating process. We compared characterizing the superconducting domain by assuming constant (field-independent) critical current density $J_{\text{c}}$ as well as a critical current density dependent on the strength and direction of the local magnetic field $J_{\text{c}}(B_{\text{x}},B_{\text{y}})$ and found that using constant critical current density achieves striking agreement with experimental measurements that could not be significantly improved (compare fig.~1). Using a functional dependency of the critical current density as in~\cite{Safran2010}, which is a generalisation of the Kim theory~\cite{Kim1963}, provides very similar results and is not necessary in this case.

For the simulation of the superconducting tape we used the following parameters: a critical current of $I_{\text{c}}=\SI{284}{\ampere}$ leads to a critical current density of $J_{\text{c}}=I_{\text{c}}/(a \times d)=\SI{2.37E10}{\ampere\per\square\metre}$. The permeability $\mu$ is defined as $\mu = \mu_{\text{r}}\times{}\mu_{\text{0}}$ with the relative permeability $\mu_{\text{r}}$ unity outside the ferromagnetic domains. The relative permeability $\mu_{\text{r}}$ inside the ferromagnetic domains consisting of electroplated nickel was determined by finding a fitting function to the data published in~\cite{Gomory2009} as follows:
\begin{equation}
	\mu_{\text{r}}(H) = -75.5 \times \{\arctan{\left( 0.00017 \times H \right)}\}^{0.9924}\\ + 119.19
	\label{eqn:mur1}
\end{equation}
where H is the magnitude of the magnetic field. The loss function is defined as the sum of the integrals over the area of the ferromagnetic domains:
\begin{equation}
	Q_{\text{FM}} = \sum_{i=1}^{2}\int_{S_{\text{FM}}} f_{\text{FM}}\ d S_{\text{FM}_{i}}
	\label{eqn:fmloss1}
\end{equation}
\begin{equation}
	f_{\text{FM}} =
	\begin{cases}
		Q_{\text{sat}} \times \left(\frac{B}{B_{\text{sat}}}\right)^2 & B \leq B_{\text{sat}} \\
		Q_{\text{sat}} & B > B_{\text{sat}}
	\end{cases}
	\label{eqn:fmloss2}
\end{equation}
with the magnetic flux density $B=\mu_{\text{0}}\times{}\mu_{\text{r}}\times{}H$, the saturation flux density $B_{\text{sat}}=\SI{500}{\milli\tesla}$ and the saturation loss $Q_{\text{sat}}=\SI{2.75}{\mega\joule\per\cubic\metre}$.  

\begin{figure}[htb]
	\includegraphics[scale=0.95]{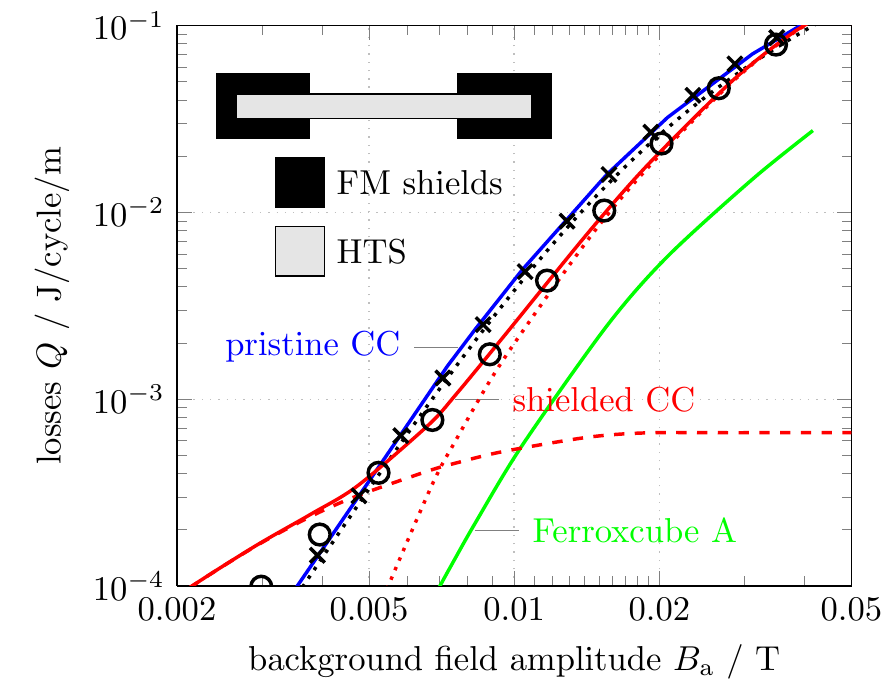}
	\caption[Hysteretic losses measured in applied oscillating magnetic background fields]{Measurements and simulations of hysteretic losses due to the application of oscillating magnetic fields; plotted are the Brandt (see Ref. 27) solution for a pristine CC (black dotted curve), the numerical simulations (solid blue) and the measurements (black crosses); also plotted are the measurements of the shielded coated conductor (black circles), the numerical simulations (solid red) and fraction contributed by the losses in the ferromagnetic material (dashed red) as well as in the superconductor (dotted red). Using Ferroxcube A (solid green) instead of nickel, the total hysteretic losses should be much lower due to its higher permeability (in these simulations, we used a constant permeability of $\mu_{\text{r}}=1400$) and lower loss per cycle ($\SI{40}{\joule\per\cubic\metre}$ instead of $\SI{2.75}{\mega\joule\per\cubic\metre}$).}
	\label{img:field1}
\end{figure}

\begin{figure}[htb]
	\includegraphics[scale=0.95]{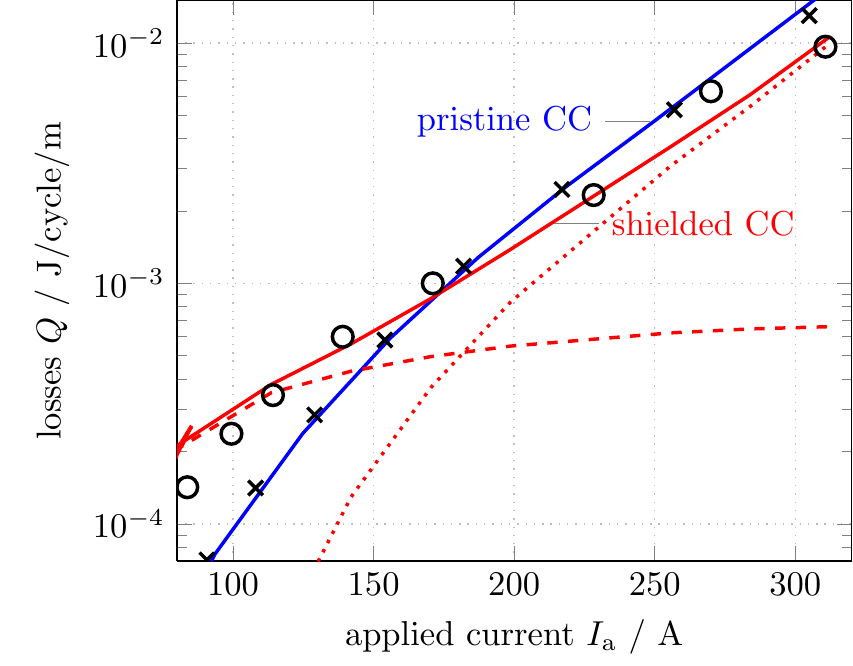}
	\caption[Hysteretic losses measured in applied oscillating transport current]{Measurements and simulations of hysteretic losses due to the application of a transport current; plotted are the measurements of a pristine coated conductor (black crosses), the simulations (solid blue), the measurements of a ferromagnetically shielded conductor (black circles) and simulations (solid red), as well as the fraction contributed by the losses in the ferromagnetic material (dashed red) as well as in the superconductor (dotted red).}
	\label{img:appliedcurrent1}
\end{figure}

Having ascertained that the coating process itself did not diminish the samples' critical current, we measured the hysteretic losses of both pristine and shielded samples when subjected to oscillating magnetic fields of varying frequency and amplitude. The samples were measured at \SI{18}{\hertz}, \SI{40}{\hertz} and \SI{60}{\hertz} and we did not observe frequency dependence of the losses per cycle which rules out eddy currents as their cause. For clarity reasons, when enough data is available, we will only show one set of measurements in the plots, namely the measurements at \SI{40}{\hertz} in the background field, and measurements at \SI{24}{\hertz} for the pristine sample and \SI{72}{\hertz} as well as \SI{1.2}{\hertz} for the shielded sample in the transport current measurements (compare fig.~1 and 2). The agreement of simulation and experiment when applying background field is generally excellent (compare fig.~1), the observed discrepancy towards very low field values is probably due to our being unable to perfectly control the ferromagnetic coats' geometry. The simulations for applied transport currents fit measured data equally well (compare fig.~2). Above roughly \SI{170}{\ampere}, the hysteretic losses in the heterostructure are reduced as compared to those in the pristine superconductor. Much as is the case for applied background fields above \SI{5}{\milli\tesla}, an effective reduction of hysteretic loss in the heterostructure assembly is only observed above certain fill rates as the reductions in hysteretic loss in the superconductor are counteracted by additional losses in  the ferromagnetic material. We would like to point out that the total hysteretic losses at low to medium fill rates are dominated by the losses in the ferromagnetic material. By modifying the coverage and thickness of the ferromagnetic coating or using different ferromagnetic material, the shield performance can be tuned for specific load scenarios.

Generally speaking, the hysteretic superconductor losses are decreased as relative permeability $\mu_{\text{r}}$ increases, resulting in a downward y-axis-shift of the curve representing the superconductor losses. This same effect is also achieved by applying thicker ferromagnetic shields although naturally at the expense of higher losses in the ferromagnetic parts of the assembly due to its higher volume. The level of saturation of the curve showing ferromagnetic losses is determined by the loss function of the ferromagnetic material. At low fields, the total losses are almost solely governed by the contribution of the hysteretic ferromagnetic losses. At low to medium fields, these can be significantly reduced by lessening the coverage of the ferromagnetic shields, thereby downsizing the volume of the ferromagnetic and hence the losses. This leads to diminished shielding at high fields however.

\begin{figure}[htb]
	\includegraphics[scale=0.95]{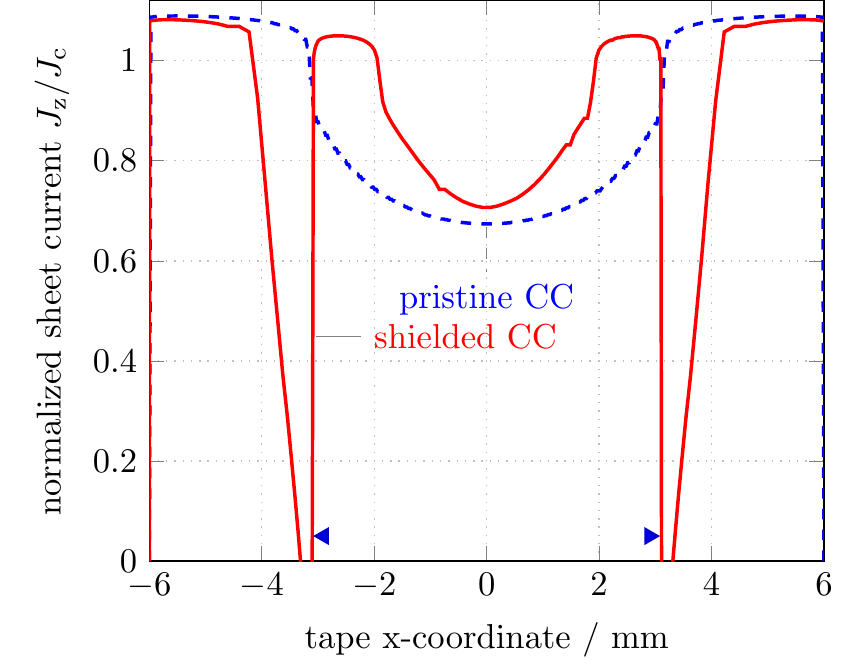}
	\caption[Comparisson of the current profiles at maximum applied current]{Comparison of the current profiles at maximum applied current for $I_{\text{applied}}/I_{\text{c}}=0.9$; current profile of the pristine sample (dashed blue) and the ferromagnetically shielded sample (solid red), the triangles mark the lateral edges of the uncovered tape; beyond, the shields extend towards the tape's edge.}
	\label{img:currentprofile1}
\end{figure}

Apart from appraising the hysteretic losses, we also investigated the current distribution in coated conductors. In shielded coated conductors, when looking at the sheet current profile of a sample with applied transport current (compare fig.~3), the observation can be made that critical currents flow in regions previously not showing them. Also, there is a dent in the current profile at the positions where the shields end and the uncovered tape begins. 

We were able to present the first comprehensive measurements of hysteretic shielding on coated conductors. Even with nickel, a material not ideally suited for ferromagnetic shielding applications, the losses at an applied field of \SI{10}{\milli\tesla} were reduced by over \SI{40}{\percent} while the loss reduction at an applied current of \SI{300}{\ampere} is over \SI{35}{\percent}. In order to obtain optimal shielding performance, the geometry needs to be tuned for a given application and load profile. The observed reduction of hysteretic losses is not as high as could be deducted from the simulations described in~\cite{Gomory2010}, however this is due to the geometry and material parameters chosen in this setup. The results shown in fig.~16 of~\cite{Gomory2010} were obtained by using a loss function with saturation value lower than that of Ni and have been corrected since~\cite{Gomory2013}. It is important to stress the fact that while the addition of ferromagnetic shields seems to be detrimental to the performance of the superconductor with applied transport current in the geometry chosen in~\cite{Gomory2010}, this is probably due to the specific dimensions of the ferromagnetic shields.

It would be highly interesting to use materials with elevated magnetic permeabilities, low loss functions and low saturation field densities. When referring to eq.~\ref{eqn:mur1} it is obvious that nickel's initial relative magnetic permeability of about $\mu_{\text{r}}=120$ which is low to begin with decreases swiftly towards higher applied fields. It is decreased by more than \SI{50}{\percent} at fields as low as \SI{10}{\milli\tesla}. Other materials should be much better suited to shield magnetic flux entry into the superconducting volume. In want of a complete dataset for Ferroxcube A, we used nickel's loss function adjusting the loss per cycle as well as the saturation flux density and used a constant permeability of $\mu_{\text{r}}=1400$ with the boundary condition of applied oscillating background fields. Exemplary simulations with the Ferroxcube A as the material for the ferromagnetic shields shown in fig.~1 demonstrate the high potential of using high permeability, low loss ferromagnetic materials. The saturation level of the hysteretic ferromagnetic losses depends foremost on the saturation field density and the hysteresis loss per cycle, so an ideal material exhibits low saturation field densities and hysteresis loss per cycle as well as high permeability. We looked for materials with magnetic properties better suited to shielding applications. Some examples are to be found in tab.~\ref{tab:magmat1}.

\begin{table*}[!htb]
	\begin{tabular}{llll}
		\toprule
		\multicolumn{4}{c}{Typical properties of several soft magnetic materials at room temperature}\\
		\midrule
		\parbox[t]{2cm}{\begin{flushleft}Material\end{flushleft}} & \parbox[t]{3.0cm}{\begin{flushleft}Initial Relative\\Permeability $\mu_{\text{r}}$\end{flushleft}} & \parbox[t]{3.0cm}{\begin{flushleft}Saturation Flux\\Density $B_{\text{s}}$ / \si{\tesla}\end{flushleft}} & \parbox[t]{3.0cm}{\begin{flushleft}Hysteresis Loss\\per Cycle / \si{\joule\per\cubic\metre}\end{flushleft}}\\
		\midrule
		Commercial iron ingot & 150 & 2.14 & 270\\
		Silicon-iron (oriented) & 1400 & 2.01 & 40 \\
		45 Permalloy & 2500 & 1.60 & 120\\
		*Ferroxcube A & 1400 & 0.33 & ~40\\
		*Ferroxcube B & 650 & 0.36 & ~35\\
		Sendust & 35000 & 1.0 & 18\\
		Hiperco & $>$3000 & 2.4 & 200\\
		Permendur & 700 & 2.45 & 300\\
		Cold-rolled Si-Steel & 1500 & 2.0 & 44\\
		78.5 Permalloy & 10000 & 1.07 & 20\\
		3.8-78.5 Cr-Permalloy & 12000 & 0.80 & 20\\
		3.8-78.5 Mo-Permalloy & 20000 & 0.85 & 20\\
		*Supermalloy & 100000 & 0.8 & 20\\
		Hipernik & 4500 & 1.6 & 10\\
		*TDK PE 22 & 1800 & 0.51 & 3.16\\
		*TDK PE 90 & 2200 & 0.53 & 2.4\\
		\bottomrule
	\end{tabular}
	\caption{Table I: Properties for ferromagnetic materials at room temperature adapted from Ref. 25, p. 822 and Ref. 26, pp. 204, 214, 231, 236 and from a material data sheet on PE 22 \& PE 90 (v.D07EA2 2010.05.17) by TDK Corporation, Tokyo, Japan. A comprehensive study at cryogenic temperatures would be most interesting. Asterisks mark especially interesting materials.}
	\label{tab:magmat1}
\end{table*}

We are certain that the ferromagnetic shielding can be applied to an array of applications and the results shown in both fig.~1 as well as fig.~2 are promising. Further research should investigate different ferromagnetic materials' potential for shielding applications in order to appraise its full potential. More data is needed on the magnetic properties of said materials. The significant reduction of hysteretic losses should provide ample motivation. Applications where these musings could prove beneficial are high current cable designs such as~\cite{Zermeno2013,Takayasu2012a,VanderLaan2013}.

This work has been parly supported by the Helmholtz-University Young Investigator Group Grant VH-NG-617.

\end{document}